\documentclass[]{scrartcl}

\usepackage{moreverb,url}

\usepackage[colorlinks,bookmarksopen,bookmarksnumbered,citecolor=red,urlcolor=red]{hyperref}

\usepackage{algorithm}
\usepackage{algcompatible}

\usepackage{caption}
\usepackage[end]{algpseudocode}

\usepackage{pgfplots}
\usepackage{tablefootnote}
\usepackage{longtable}
\usepackage{tabularx}
\usepackage{booktabs}
\usepackage{multirow}

\usepackage{pdflscape}

\usepackage{authblk}

\usepackage{url}

\usepackage{natbib}

\usepackage{amsmath}
\usepackage{amssymb}

\usepackage[l3]{csvsimple}

\title{Beyond Bars: Distribution of Edit Operations in Historical Prints}

\author[1]{Adrian Nachtwey}
\author[2]{Fabian C. Moss}
\author[1]{Anna Viktoria Katrin Plaksin}

\affil[1]{KreativInstitut.OWL, Paderborn University, Paderborn, Germany}
\affil[2]{Institut f\"ur Musikforschung, Julius-Maximilians-Universit\"at W\"urzburg, W\"urzburg, Germany}

\begin{document}
\maketitle

\begin{abstract}

In this paper, we present a method for conducting comparative corpus studies in musicology that reduces the time-consuming digitization process. Instead of encoding whole corpora of musical sources, we suggest sampling bars from these sources. We address the challenge of selecting representative samples and evaluate three different sampling methods. We used Beethoven's Bagatelles Op. 33 as a case study to find the method that works best in finding samples representative with respect to differences. 
We believe that this approach offers significant value to musicological research by enabling large-scale analyses and thereby statistically sound results. Moreover, we believe our work to be a valuable step toward understanding nineteenth-century editorial practices and enriching the field of scholarly editing of historical musical works.

\end{abstract}

\section{Introduction}
\label{intro}

Until modern forms of music dissemination such as audio recordings were developed, music lovers could only get to know new music by playing it at home using prints available to them, or by going to concerts. As concerts were not affordable or accessible for everyone, prints heavily influenced the shape in which the audience got to know a new piece of music and thus are crucial for its transmission (see \cite{Lewis2024}).
Regardless, music publishers and the products of their work, musical prints, have been mostly ignored by musicological research. Exceptions are prints authorized by the composer. Of interest are mostly first editions and rarely reprints when the composer was involved in their creation. This has been acknowledged only recently by the project ``Das Handwerk des Verlegers" (``The publisher's craft") at the Beethoven-Haus Bonn (2021—2024)\footnote{See \url{https://www.beethoven.de/de/g/Handwerk-des-Verlegers}.} the aim of which was to also examine non-authorized prints of Beethoven's works impartially (for research on editors, see also \cite{Beer2000} and \cite{Beer2020}).

Our research follows this approach.
We want to investigate differences between various editions of Beethoven's piano music. Other than the ``Handwerk" project we do not want to focus on selected publishers known to have worked with Beethoven, but expand the scope of the research. Also, we do not want to mainly do qualitative analysis of different states of printing plates. Instead, we want to focus on quantitative analysis. In a first step, we want to compare the editions, find differences and count them. In this way, we want to shed light on the relationships between different editions as well as the shape in which the broad interested public got to know these works.
Also, we want to examine the hypothesis, that in 19\textsuperscript{th}-century German-speaking countries a development took place, that led to a consolidation of the text of a work, inspired by \cite{Goehr1994}. Goehr argues that the concept of the Musical Work emerged around 1800 and gained more and more impact on the musical life throughout the nineteenth century. Given this to be true, we believe that the impact should also be evident in the practice of music editing in this century. We believe that this could lead the editors to aim to publish a musical text in a shape that we today call the ``Urtext''. This, then, could lead to a convergence of the musical text in various editions because they all aim to recreate the same text.

Like most musicological corpus studies, we too face the problem of acquiring our data (see \cite{Lewis2023}). As Optical Music Recognition (or \textit{OMR}) techniques are, despite recent advances (see \cite{Mayer2024}, \cite{Rios-Vila2023}, \cite{Rios-Vila2024}), limited regarding the quality of their output or their field of application (for a general state of the art see \cite{calvo-zaragoza-2023}), they are no option for our and the work of other projects to automatically encode the scores and so musical texts have to be encoded manually in time-consuming work. 
Therefore, musicologists often rely on already existing corpora. 

When creating corpora, however, researchers are confronted with issues of representation and representativeness, depending on the research question at hand. For example, \cite{London2013} tries to build a corpus representative of listeners' consumption of classical music or `a classical music demographic' (p. 69). Several sources are analyzed, and the author considers it biased that ``unpopular'' classical music pieces are included in lists of ``essential'' recordings based on their compositional value ascribed by musicologists or other critics (ibid.).

\cite{Shea2024}, on the other hand, try to create corpora not biased by the self-reinforcement of already popular music,
the popularity of the composer or musician and by conscious or unconscious racism, sexism, and other forms of structural discrimination. This bias extends to popular media representations, such as the \textit{Rolling Stone Magazine} lists analyzed by \cite{Shea2024} indicating a broader structural issue. The tendency to marginalize certain ethnic groups and genders is evident not only in popular music but also in musicology, which typically centers on few composers of Western art music.
The same is true for the canon of classical works of Western art music being played in concert or recorded (see \cite{London2013}). But, as we have seen, there is not ``the one'' kind of representativeness. Depending on the aim of a study, a corpus can be representative in different ways. It can be representative in regard to the population of a country (\cite{Shea2024}) or in regard to the music a 21\textsuperscript{st}-century listener is most familiar with (\cite{London2013}). 

The contrast between the two cited works is an example for this difficulty: The two studies examine music from different points of view and thus create two different corpora, both representative in their own right. Because of the differing questions, each corpus is only usable for the study for which it has been created and not for the other.

There have been efforts made to create corpora for further work, for example \cite{burgoyne2011}, \cite{devaney2015}, \cite{hentschel2025} and \cite{neuwirth2018} to name a few. The problem with preexisting corpora is that they either have to fit another research question by chance or researchers have to make their questions fit the corpora, which makes them only useful to a handful of projects. Additionally, this again reinforces the use of music by already popular composers and musicians, preventing a more diverse musicological research.

The main challenge for the present work is that the number of relevant prints of Beethoven's piano music is huge; for example, there are on average more than 50 prints per sonata in the archives of the Beethoven-Haus Bonn alone. \cite{Saccomano2024} propose sampling as a solution to this kind of problem, but the aim of the authors is to present methods to capture information about the sampled parts and the gaps in an encoding to make the data useful for further research. They do not address the sampling process itself. This is what we deal with in this paper.

To avoid the mentioned problems with using pre-existing corpora and fitting our research to them, we decided to go another way: we do not encode the whole prints for the comparisons but only samples. This means, we want to include as many editions as possible of Beethoven's piano music, but we only want to use a small part of all the bars of this music. In this way, we minimize the workload in encoding the different editions and at the same time do not rely on already encoded editions. 
But drawing the samples raised delicate questions: How should the samples be drawn so that they are representative of the whole piece in regards to differences between various editions? Are differences equally distributed throughout the prints so that a random selection of bars suffices? Or are there some factors that influence the distribution of differences that must be taken into account in the sampling process? Here, we present the results of a case study evaluating three different methods for sampling bars from musical pieces.

\section{Data}

As a case study, we used ``small'' pieces by Beethoven, namely the seven Bagatelles Op. 33. We chose these pieces, i.e. short pieces, because they can easily be encoded in full length.

Our test corpus consists of the Bagatelles Op. 33 in the six following editions. The publication dates are rough estimations as given in the description of the editions in the Beethoven-Haus catalog or the Petrucci Library, respectively.
\begin{enumerate}
\item the first print by \textit{Bureau d'Arts [sic!] et d'Industrie} (Vienna 1803, in the following: ``BdA'')\footnote{\url{https://www.beethoven.de/de/media/view/4956805541134336/scan/0}}, 
\item prints by Zulehner (Mainz 1808)\footnote{\url{https://www.beethoven.de/de/media/view/4697437075668992/scan/0}},
\item André (Offenbach a.M. 1825)\footnote{\url{https://www.beethoven.de/de/media/view/4908887509565440/scan/0}},
\item Schott (Mainz 1826)\footnote{\url{https://www.beethoven.de/de/media/view/5114772269826048/scan/0}},
\item Haslinger (Vienna 1845, ``Hasl.'')\footnote{\url{https://www.beethoven.de/de/media/view/5673265322262528/scan/0}} and 
\item Breit\-kopf \& Härtel (Leipzig 1864, ``Breitk.'').\footnote{\url{https://imslp.org/wiki/Special:ReverseLookup/58117}}

\end{enumerate}
We used the MEI encoding of this Op. 33 by \cite{Goebl2021_MEIEncodingSieben}\footnote{\url{https://github.com/trompamusic-encodings/Beethoven_Op33_BreitkopfHaertel.git}}  as a basis for our work. 

\section{Method}
\label{sec:method}

\subsection{Sampling}
The data generation model in this study can be described as follows. Suppose $I$ opaque jars that contain marbles of different colors. Each jar contains only marbles of one color. Let the number of marbles in a jar be $n_i$ and $N=\sum_{i}n_i$ the total number of marbles. Each jar may contain a different number of marbles. Now, we randomly draw $\tilde{n}_i$ marbles from jar $i$ until we get a sample of some size $S$. 
The condition of a sample being representative of the jars now means that the distribution of marbles of different colors in $S$ is the same as in the jars, so the ratios $\frac{n_i}{N}$ and $\frac{\tilde{n}_i}{S}$ are approximately equal. 

\begin{figure*}[h]
\begin{center}
\begin{minipage}[t]{0.49\textwidth}
\includegraphics[width=\textwidth]{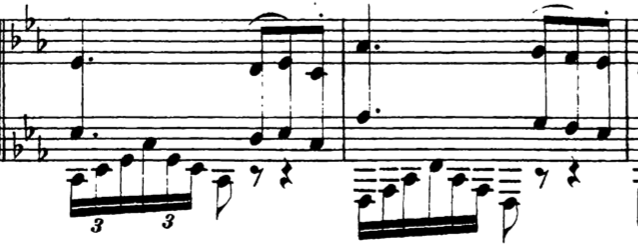}
\caption*{Edition by Breitkopf \& Härtel.}
\end{minipage}
\hfill
\begin{minipage}[t]{0.45\textwidth}
\includegraphics[width=\textwidth]{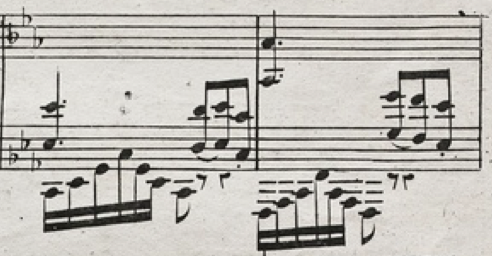}
\caption*{Edition by Schott.}
\end{minipage}
\caption{Example for differences between the editions of the 5th Bagatelle, bars 25 and 26, by Breitkopf \& Härtel and Schott. They differ in the placement of the right hand melody, of the articulations (slur and staccato) of this melody and the numbers to indicate triplets. Also, different symbols for quarter rests are used.}
\label{fig:editions}
\end{center}
\end{figure*}

In this analogy, the colors represent the different kinds of differences found between two musical prints. An example of various differences can be seen in Fig. \ref{fig:editions}. In this example, the allocation of the notes on the two staffs is different, as is the placement of slurs and staccato dots (above the notes in the example on the left, below on the right). Also the numbers used to indicate triplets are omitted in the Schott edition and the two editions use different symbols to represent quarter rests.  The number of marbles represents the number of times, a specific kind of difference can be found in the comparison of two prints.

We now have several options for drawing the samples, for example, drawing equally from all jars assuming that differences are evenly distributed.
If for some reason we think that one color (or kind of differences) occurs more often (i.e., there are more marbles of one color than of the others), we could bias our sampling process so that it is more likely to draw from one of the jars than from another. This is what we did in this case study where the distribution of marbles in the jars represents the distribution of differences in the scores and the three sampling methods under investigation represent three assumptions about this distribution.

\subsection{Sampling algorithms}
\label{subsec:algos}
Using three algorithms, we drew the samples. This means, we selected bars from the full encodings. We used bars as the basic unit because the comparison with \textsf{musicdiff} (see below) only works on valid MEI-files. Also, the notation of music only makes sense in context. A single note without at least a clef and key signature is meaningless. Thus, differences can not be correctly identified without considering their surroundings.

All scripts are written in Python-code. For mathematical operations and data handling the \textsf{numpy} und \textsf{pandas} libraries were used. \textsf{musicdiff} depends on the \textsf{music21} library as well as the \textsf{converter21} library.
The algorithms take as an argument a Python dictionary called ``measures''. It contains as keys the bar numbers as given as \texttt{@n}-attributes in the MEI-files and as values the number of musical elements inside the bars. These elements are not to be confused with the MEI elements used to encode the music.

To count the elements and copy the selected bars to new files we especially used the Python module \texttt{xml.etree.ElementTree}. Since MEI is an XML-based encoding, it can be parsed by \texttt{ElementTree}. The musical elements to be taken into account were given as a list. All relevant scripts and full-length encodings are available as a Docker Image on GitHub.\footnote{see \url{https://github.com/CorpusBeethoviensis/beethoven-diff-docker.git}. For a complete documentation, see \url{https://deepwiki.com/CorpusBeethoviensis/Docker/1-overview}.} 

\subsubsection{Algorithm 1: Random selection (randSel)}
\label{subsubsec:random}
First, we used an algorithm that randomly selects bars from the piece without any knowledge about the contents of the bars (see Algorithm \ref{algo:random}, p. \pageref{algo:random}). This means that the algorithm randomly draws numbers out of a list of numbers ranging from 0 (incomplete pickup beats where encoded with \texttt{@n=0}) to the total number of bars in the piece.

\subsubsection{Algorithm 2: Selection based on bar number and element count (barElCount)}
\label{subsubsec:elembars}
The second algorithm based the selection on a second information, namely the number of musical elements inside the bars. With this algorithm we still controlled the number of bars that were to be selected but we used as a second requirement that the selected bars should contain a matching number of notational elements, i.e. 10\% of the elements in the whole piece (see Algorithm \ref{algo:mae}, p. \pageref{algo:mae}). For more flexibility we did not set the number of elements as a fixed value to be matched exactly but allowed the samples to be 5\% smaller or larger than the exact number. If 10\% of the elements were 100, we would allow 95-105 elements.

\subsubsection{Algorithm 3: Selection based only on elements (onlyEl)}
\label{subsubsec:onlyEl}
The third algorithm selects bars only taking into account the number of musical elements. The percentage of selected bars is not preset. This means that the number of bars may be (considerably) smaller or bigger than the ~10\% of musical elements (see Algorithm \ref{algo:elem}, p. \pageref{algo:elem}). Here, again, the number of elements may vary between 95 and 105\% of the required number.

\subsection{Differences in music encodings}
We created six files with Goebl's encoding and modified it so that each file matches one of the six editions under investigation.  
We then compared the editions to each other using the Python-Tool \textsf{musicdiff}\footnote{The documentation can be found at \url{https://github.com/gregchapman-dev/musicdiff}, see \cite{Foscarin2019}.}, obtaining the number of edit operations $\delta_{e,e'}$ (with $e,e' \in \{1,6\}$ denoting the editions and $e\neq e'$) necessary to convert one edition into the other. \textsf{musicdiff} produces text files that contain a list of edit operations, ordered chronologically. Each edit operation is listed with the location of its appearance in the format \texttt{@@ measure n, staff 1/2, beat x@@}. We simply counted the number of these lines as the number of edit operations. The results can be seen in the Appendix, Tables \ref{tab:full_results-1}—\ref{tab:full_results-5}, p. \pageref{tab:full_results-1}.

After this, we drew the samples. We always used the first print by \textit{Bureau d'Arts et d'Industrie} as the basis for our samples. For thoughts on biases of using a particular edition as a basis, see \ref{subsec:samples}. We repeated this sampling process 10,000 times per Bagatelle and algorithm. This results in 30,000 samples per Bagatelle, 210,000 in total. For each sample, we created new files. The first of the seven Bagatelles contains 97 bars so with the first two algorithms each sample contained 10 bars. For each of the editions we thus created 10,000 valid MEI files that only contain the 10 sampled bars each for the \textit{randSel} and the \textit{barElCount} algorithm and an arbitrary number of bars for the \textit{onlyEl} algorithm. These sample files were then compared pairwise using \textsf{musicdiff}. 
With six editions under consideration, we get 15 comparisons per sample. This results in 450,000 comparisons per Bagatelle. For each sample $s$ we obtain the number of differences between each pair of editions, $e$ and $e'$, $\delta_{e,e'}(s)$. We tested if the samples represent the whole encoding by comparing the number of differences in the samples to the theoretical expectation value $\mu_{e,e'}$. 

This means, in other words, a sample of size $x\%$ of the bars would contain $x\%$ of the edit operations compared to the whole piece, or $\delta_{e,e'}(s)\approx~x\delta_{e,e'}$, where $x \in {[0,1]}$. 
From the total number of differences we calculated the theoretical mean, i.e. 10\% of the total, which is the theoretical expectation value $\mu_{e,e'}$, the number of differences the samples should contain in order to represent the full pieces.

Unfortunately, we had to exclude the Bagatelles Nos. 6 and 7 from the analysis. In the case of No. 7 the reason for this is that the comparisons failed because the Breit\-kopf \& Härtel edition has more bars than all the others: bar 86 is part of the first and second ending of a repetition only in this print. All the other prints start with the endings in bar 87. For this reason, all the following bars have differing \texttt{@n}-attributes in the Breitkopf edition. Since the files with the sampled bars are created automatically, we would compare bars to each other that do not contain the same music. This is not reasonable much like manually creating 30,000 sample files is not. 

For the 6\textsuperscript{th} Bagatelle and almost 3,000 samples one or more comparisons failed. The reason is not obvious to us. We tried to reproduce the error, and in some cases the comparisons still failed; in others, they produced the desired output. This leads us to assume that the problem is not exclusively with our data, but in parts with \textsf{musicdiff} or our comparison pipeline. Since there are so many samples for which comparisons have failed, we had to refrain from doing the comparisons manually.

Of course, it would also be possible to just extract the distribution of differences using the comparisons of the full-length encodings. But since we are interested in the performance of three potential sampling algorithms, this does not suffice. Even if we knew this distribution, we would not know if the algorithms perform well on real data. Especially we would not know the general accuracy with which the algorithms draw samples. We can only know this through repeated trying because only then we are able to calculate the mean accuracy as well as the standard deviations.

To be sure that the results are not distorted by one or more identical samples we checked the drawn samples. Especially for the algorithm described in \ref{subsubsec:elembars} it was not obvious if there were enough possible combinations.\footnote{For the completely random selection there exist 1,198,774,720 combinations for 70 bars and a selection of seven bars, for 100 bars and a sample size of 10 there are 17,310,309,456,440 possibilities.}
But in only two cases, both for Bagatelle No. 5, with the \textit{randSel} and the \textit{onlyEl} algorithms, one sample has been drawn twice. In all the other cases, we obtained 10,000 different samples. With this in mind, our results are not influenced by counting the same comparisons more than once. 

\begin{figure}[h]
\begin{center}
\includegraphics[width=\textwidth]{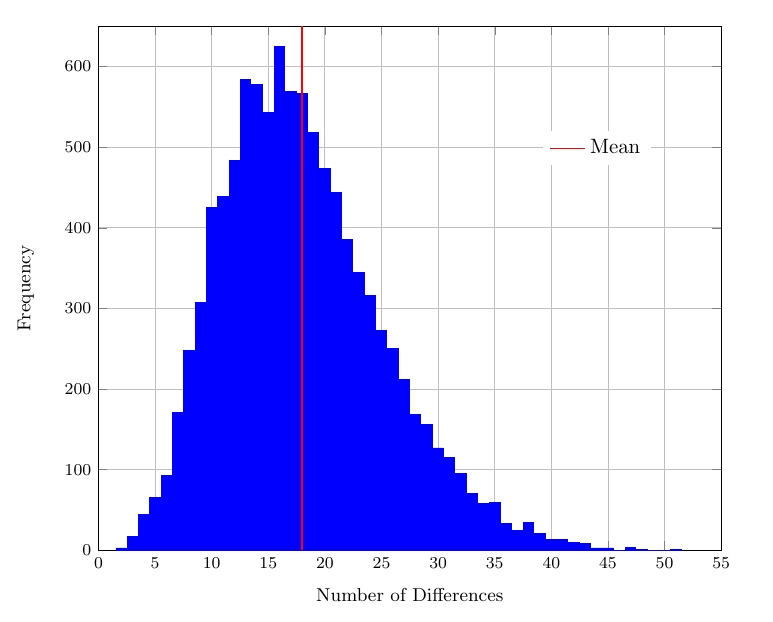}
\end{center}
\caption{Example plot of the results, Algorithm 3, Bagatelle 1. The red line shows the empirical mean $\bar{\delta}_{e,e'}$.}
\label{fig:result}
\end{figure}

For each pair of editions we visualized the results so that the x-axis represents the number of edit operations and the y-axis the frequency with which this particular number of edit operations occurs (see Figure \ref{fig:result}). We obtain distributions similar to the normal distribution skewed to the right.

We know the expected theoretical mean of differences ($\mu_{e,e'} = 0.1*\delta_{e,e'}$) we want to obtain using our three algorithms. We now compare the results of the sampling process, the empirical means $\bar{\delta}_{e,e'}$, to the theoretical means.

We calculated the difference between the expected number of differences for each Bagatelle and each pair of editions and the mean of the algorithms. With this, we calculate the average of this difference 
\begin{equation}
\Delta = \binom{6}{2}^{-1}\sum_{\substack{e,e'=1\\e\neq e'}}^{6}|\mu_{e,e'}-\bar{\delta}_{e,e'}|,
\end{equation}
as well as the mean of the standard deviation (see Table \ref{tab:results}).

\begin{table*}
\caption{Average deviation of the empirical means ($\Delta$) and the average standard deviation ($\sigma$) of the three algorithms per Bagatelle. The values were rounded to two decimal places.}
\label{tab:results}
\centering
\begin{tabular}{lcccccc}
\toprule
\textbf{No.}& $\Delta_{randSel}$ & $\Delta_{barElCount}$ & $\Delta_{onlyEl}$ & $\sigma_{randSel}$ & $\sigma_{barElCount}$ & $\sigma_{onlyEl}$\\ % \hline
\midrule

 1 & 1.95 & 1.87 & 3.32 & 9.62 & 9.46 & 9.24 \\
 2 & 2.67 & 2.75 & 5.19 & 8.97 & 8.91 & 9.68 \\
 3 & 3.22 & 3.25 & 4.24 & 9.47 & 9.17 & 9.00 \\
 4 & 3.92 & 3.71 & 4.15 & 7.57 & 7.52 & 7.67 \\
 5 & 2.54 & 2.80 & 2.96 & 14.63 & 14.96 & 14.97 \\ 
 \midrule
$\varnothing$&2.86&2.88&3.97&10.00&10.00&10.11\\
\bottomrule
\end{tabular}
\end{table*}

We use the results to evaluate the performance of the algorithms. The main criterion is the $\Delta$ which should be as small as possible. In addition to that we take into account the standard deviation. If two or all three algorithms generate similar $\Delta$ we choose the one with the smallest standard deviation. Depending on the results it will be necessary to weigh a smaller $\Delta$ against a smaller standard deviation and \textit{vice versa}.

\section{Results}
\label{sec:results}
All three algorithms perform very well. They only differ slightly.
As Table \ref{tab:results} shows, the first algorithm (\textit{randSel}), choosing bars completely randomly, differs the least from the theoretical mean; on average it deviates by 2.86 differences. The average standard deviation is 10.00 differences.

The second algorithm (\textit{barElCount}) deviates by 2.88 differences with a standard deviation of 10.00 differences, too. 
The third algorithm differs the most with a deviation of the mean of 3.97 differences and with a standard deviation of 10.11 differences. Due to the smallest difference of the theoretical and empirical mean combined with the smallest standard deviation, we believe the \textit{randSel} algorithm to perform best, closely followed by the \textit{barElCount} algorithm. Table \ref{tab:results} shows that the difference between these two algorithms is not only small in their mean deviation but that for three Bagatelles, Nos. 2, 3 and 5 the \textit{randSel} algorithm has the smaller deviation, and for the two remaining Bagatelles the \textit{barElCount} algorithm has the smaller deviation. Tables \ref{tab:full_results-1}—\ref{tab:full_results-5} show how close the results of these two algorithms are. Thus, we think the choice to be arbitrary. As a decisive reason to choose the \textit{randSel} algorithm over the \textit{barElCount} algorithm we took into account the complexity of the algorithms and consider the random selection to be computationally less costly.

\section{Discussion}
\subsection{A closer look: Bagatelles Nos. 3 and 5}
It is worth noticing that in almost all cases the empirical and theoretical mean are consistent within the standard deviation of the empirical values. There are only three comparison pairs where all the algorithms differ from the theoretical values: for Bagatelle No. 3 the comparisons of the edition by Zulehner with Hasl. and Breitk., respectively; for Bagatelle No. 4, again the comparison of Zulehner's edition with the one by Hasl. In the case of the \textit{onlyEl} algorithm, there is a fourth comparison, Bagatelle No. 5 comparing editions by Zulehner and André.

In the case of Bagatelle 3 the samples overestimate the number of differences (see table \ref{tab:full_results-3}). The reason seems to be that the comparison of the full files did not produce the right output. There are many bars that have not been compared in detail. Instead, they have been recognized as missing. This leads to a much lower number of differences in the comparison of the complete files and thus to a much lower $\mu_{e,e'}$. 

There is another reason for a closer look at Bagatelles 3 and 5. The results of the algorithms mostly underestimate the ``real'' number of differences.
Exceptions are the following:

\begin{itemize}
    \item Algorithm \textit{randSel}:
    \begin{itemize}
        \item Bag. 3, 4 comparisons.
        \item Bag. 5, 11 comparisons.
    \end{itemize}
    \item Algorithm \textit{barElCount}:
    \begin{itemize}
        \item Bag. 3, 6 comparisons (including all the ones from \textit{randSel} algorithm).
        \item Bag. 5, 11 comparisons (the same as for \textit{randSel} algorithm).
    \end{itemize}
    \item Algorithm \textit{onlyEl}:
    \begin{itemize}
        \item Bag. 3, 2 comparisons. 
        \item Bag. 5, 3 comparisons.
    \end{itemize}
\end{itemize}

\begin{figure*}
\begin{center}
\includegraphics[width=\textwidth]{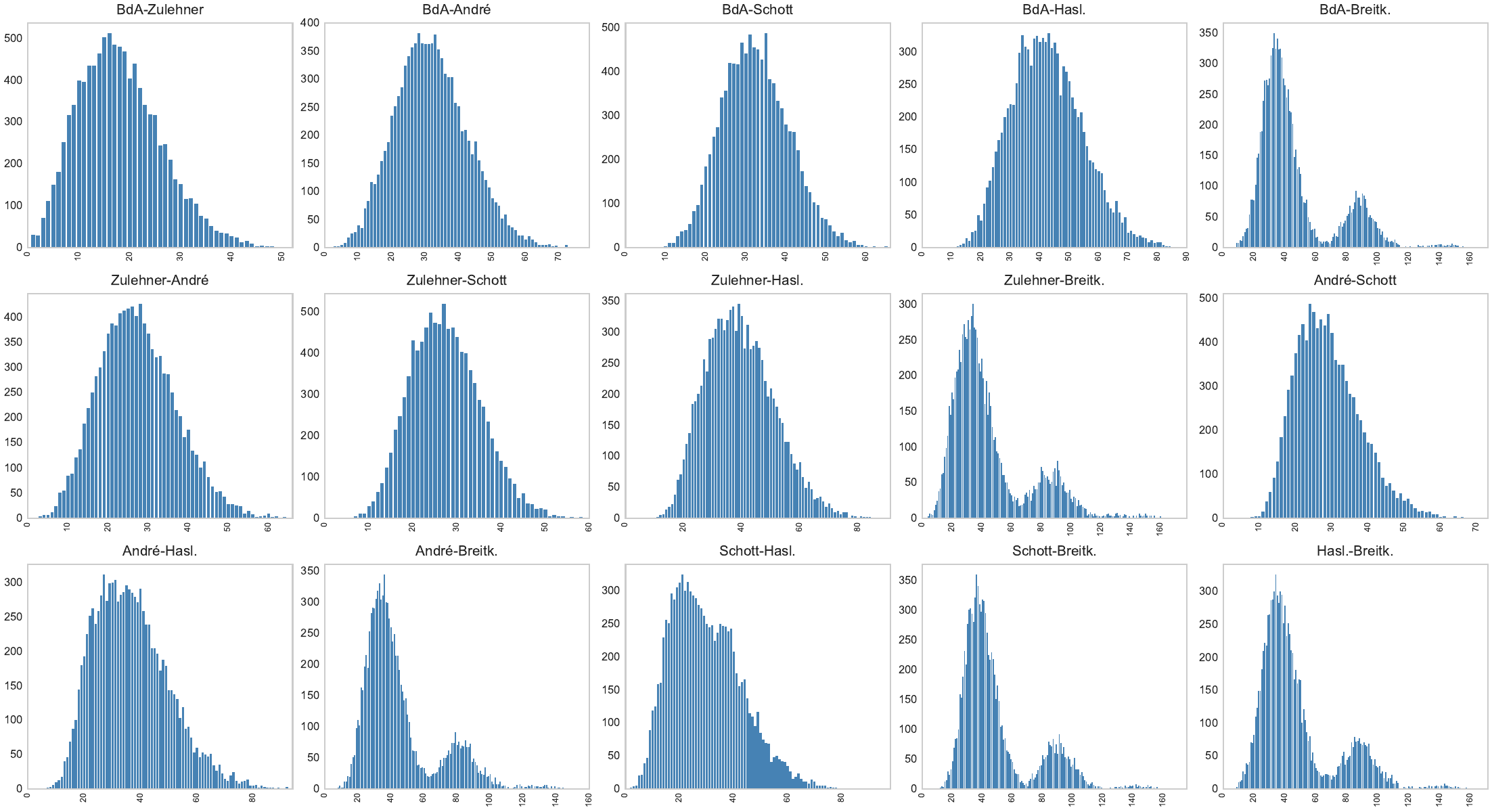}
\caption{Results of the comparison of Bagatelle No. 5 for the samples drawn with the \textit{onlyEl} algorithm.}
\label{fig:ex_5}
\end{center}
\end{figure*}

It is also interesting that the algorithms work very well even for Bagatelle No. 5. As can be seen in Figure \ref{fig:ex_5}, there are some comparisons whose results are not only not normally, but bimodal distributed. These are the comparisons involving the edition of Breitkopf \& Härtel. The results look similar for all of the algorithms. Still, the empirical means are very close to the theoretical means. Since the deviation of the data is rather large, the standard deviation is also large compared to the standard deviations for the other Bagatelles (see Table \ref{tab:results}).

\begin{figure*}
\begin{minipage}[c]{0.45\textwidth}
\includegraphics[width=\textwidth]{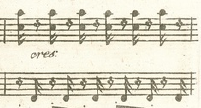}
\caption*{a) Relevant bars as printed in the first edition by Bureau des Arts.\footnote{First print in possession of the Beethoven-Haus Bonn: \url{https://www.beethoven.de/de/media/view/4956805541134336/scan/0}}}
\end{minipage}
\hfill
\begin{minipage}[]{0.45\textwidth}
\includegraphics[width=\textwidth]{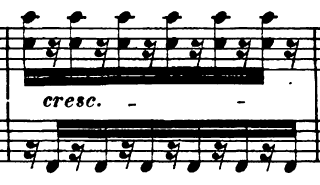}
\caption*{b) Relevant bars as printed in the edition by Breitkopf \& Härtel.\footnote{To be found at the Petrucci Library: \url{https://imslp.org/wiki/Special:ReverseLookup/58117}}}
\label{fig:ciritcal_bars_r}
\end{minipage}
\caption{Bar 63, one of the bars responsible for the results in Fig. \ref{fig:ex_5}.}
\label{fig:critical_bars}
\end{figure*}

The reason for the strange distribution of the results seems to be mainly two bars, Nos. 63 and 64. The former can be seen in Figure \ref{fig:critical_bars}, the second looks the same. Only in the edition by Breitkopf \& Härtel Figure \ref{fig:critical_bars} b), there are beams to indicate the 16th notes with the rests in between the notes somehow belonging to the beams. In all the other editions this is printed as can be seen in the example from the first edition. As a consequence, the comparison finds different beaming for every note/chord and every rest, resulting in an extremely high number of differences in these two bars. The second maximum that can be seen in some of the results in Figure \ref{fig:ex_5} probably represents all the samples containing these two bars.

In summary, despite the mentioned observations of conspicuous cases, all algorithms work well. The big differences between the theoretical and the empirical data can easily be explained and classified in their impact on this study.

\subsection{A closer look: \normalfont{onlyEl} Algorithm}
\label{subsec:samples}
As the third algorithm \textit{onlyEl} selects bars based on the number of elements in the piece, another question arose: what encoding, i.e. which edition, should we use as our basis for sampling? 
Usually, all of the editions contain the same number of bars (except for cases like the one discussed for Bagatelle No. 7, where the Breitkopf edition contained one more bar than the others), but it was clear that the number of elements in the editions must be different. 
However, analysis shows that the differences in musical elements in the six editions are marginal, as can be seen in Table \ref{tab:element_count}. 

\begin{table*}
\centering
\caption{Number of musical elements in the six editions of the Bagatelles Op. 33. The numbers highlighted in red show cases, where prints contain a strikingly high number of differences in comparison to the other prints.}
\label{tab:element_count}
\begin{tabular}{l|cccccc|rrr}
\toprule
\textbf{No.}& \textbf{BdA} & \textbf{Zulehner} & \textbf{Andre} & \textbf{Schott} & \textbf{Hasl.} & \textbf{Breitk.}& \textbf{mean} & $diff_{max}$&$diff_{rel}$\\ 
\midrule
1 &2152 & 2151 & 2124 & 2151 & 2186 & 2211&2162,5&87&0,0402\\
2 & 2163 & 2159 & 2127 & 2152 & \textcolor{red}{2453}&\textcolor{red}{2446}&2250&326&0,1449\\
3 &1311&1312&1288&1313&\textcolor{red}{1451}&1356&1338,5&163&0,1218\\
4&1483&1481&1466&1489&1491&1579&1498,17&113&0,0754\\
5&3162&3156&3102&3121&3107&3162&3135&60&0,0191\\
6&1380&1377&1364&1377&1399&1405&1383,67&41&0,0296\\
7&2299&2295&2300&2296&2369&2430&2331,5&135&0,0579\\ 
\bottomrule
\end{tabular}
\end{table*}

The numbers highlighted in red show cases where prints contain a strikingly large number of elements compared to the others, resulting in a comparatively large relative difference of more than 14 and 12\%, respectively, opposed to values smaller 8\%. The 2\textsuperscript{nd} Bagatelle is another case with different numbers of bars in the editions: in the two editions by Haslinger and Breitkopf \& Härtel a section is printed that is the repetition of an earlier section. In all the other editions this is indicated by \textit{Scherzo Da Capo Senza repetizione e dopo el trio}. 
Without the additional bars, the two editions contain 2197 (Haslinger) and 2182 (Breitkopf) elements, resulting in a maximal relative difference of about 5,5\%, which is much more aligned with the relative differences for the other Bagatelles.

\begin{comment}
\begin{table*}
\small\sf\centering
\begin{tabularx}{0.75\textwidth}{c|cccccc|ccc}
& \multicolumn{6}{c|}{Editions}&&\\\hline
No.& BdA & Zulehner & Andre & Schott & Hasl. & Breitk.& Mean & \(diff_{max}\)&\(diff_{rel}\)\\\hline

2 & 2163 & 2159 & 2127 & 2152 & \textcolor{red}{2197}&\textcolor{red}{2182}&1263,3&70&0,0554\\

\end{tabularx}
\caption{Number of relevant elements in the six editions of the \textit{Bagatellen} Op. 33 without additional bars.}
\label{tab:element_count_corr}
\end{table*}
\end{comment}

The big difference in the number of elements in the Haslinger edition of the third Bagatelle is a direct consequence of the difference between the editions. In the Haslinger print, there are many more elements that indicate articulations like \textit{legato} or \textit{staccato}. This can be seen in Table~\ref{tab:number_elements_3}. Still, because the number of musical elements in the editions only vary by percentages below 8\%, this is a first hint on the nature of differences between the editions: it seems like the number of differences occurring because of added or deleted elements might be considerably smaller than the number of differences occurring because of changed appearance of elements. This concerns, for example, changed stem directions, changed length of slurs, or changed positioning of dynamics and articulations.

\begin{table*}
\caption{Number of different elements in the Bagatelle Op. 33 No. 3. Highlighted in red are strikingly large numbers for \texttt{artic}- and \texttt{slur}-elements in the editions of Haslinger and Breitkopf \& Härtel.}
\label{tab:number_elements_3}
\centering
\begin{tabular}{lrrrrrr}
\toprule
\textbf{Element} & \textbf{BdA} & \textbf{Zulehner} & \textbf{Andre} & \textbf{Schott} & \textbf{Hasl.} & \textbf{Breitk.}\\
\midrule
\texttt{beam} & 221 & 223 & 222 & 223 & 222 & 224\\
\texttt{note} & 850 & 851 & 849 & 851 & 850 & 850\\
\texttt{rest} & 23 & 23 & 24 & 23 & 23 & 23\\
\texttt{artic} & 30 & 32 & 26 & 32 & \textcolor{red}{131} & 32\\
\texttt{tempo} &- & -& 1 & 1 & 1 & 1\\
\texttt{dynam} & 50 & 49 & 41 & 50 & 49 & 49\\
\texttt{dir} &- & 1 &1 & 1 & 1&-\\
\texttt{slur} & 51 &50 & 46 & 47 & \textcolor{red}{98} & \textcolor{red}{89}\\
\texttt{chord} & 17 & 17 & 16 & 17 & 12 & 19\\
\texttt{accid} & 54 & 51 & 51 & 53 & 52 & 54\\
\texttt{tie} & 15 & 15 & 11 & 15 & 12 & 15\\
\bottomrule
\end{tabular}
\end{table*}

With these results it shouldn't matter, which of the encodings serves as a basis for the 3\textsuperscript{rd} \textit{onlyEl} algorithm so that we again chose the first print for consistency.

All algorithms represent different assumptions about the distribution of differences between editions of musical pieces. The underlying assumption for the first algorithm (\textit{randSel}) is that the differences are distributed completely equally. This means that the probability of finding a certain number of differences is equal for every location in the score. The assumption represented by the second (\textit{barElCount}) and third (\textit{onlyEl}) algorithms is that there is a correlation between the number of differences and the number of musical elements; in other words: when the number of musical elements in a bar is comparatively high, the probability of finding differences in this bar is higher than in a bar with a small number of elements. 

These assumptions also represent different notions of the process of creating an edition of a musical work. If the number of differences is correlated with the density of musical elements, this could indicate that most differences occur because of deliberate choices by editors leading to what we will call ``propagating differences". For example, if the stem direction of one or more groups of 16th notes changes, this would be an editorial decision. Also, if the stem direction of the first note of such a group changes, this affects all following notes, hence the name ``propagating differences". Additionally, this can be a hint on the susceptibility to errors of the engraver: if the number of differences and the element density correlate one might assume that more complex parts of the score lead to more errors by the engraver.

If, however, differences are distributed randomly, one could assume that editors' choices mostly lead to local changes instead of propagating differences and that the probability of engravers making errors is independent of the complexity of the score.

The \textit{barElCount} algorithm is a compromise of both possibilities. Since it is limited in the selection process, both possible reasons for the differences are represented in its samples. First, because of the condition that a certain number of elements must be matched, the assumed correlation between differences and element density is implemented. Second, because of the condition that a certain number of bars needs to be chosen, the assumed uniform distribution of differences is implemented.
Since we believe the first hypothesis to be more convincing, we expected the \textit{onlyEl} algorithm to work best. 

This hypothesis is not supported by our findings. On the contrary: the algorithm selecting bars only based on the number of elements performs worst, the completely random selection offers the best results. The \textit{barElCount} algorithm, which combines the methods of the other two, performs equally well.

We believe this to be due to the fact that the total number of differences is composed of two factors that have varying degrees of influence: Most of the differences are local differences. This means that editors' choices mostly affect single elements like slurs or added dynamics or articulations instead of groups of elements. In addition to that, there are errors made by the engravers. It seems like the probability for engravers' errors is not correlated with the element density or complexity of the score.

Hence, the first algorithm that randomly selects bars performs best. The second algorithm still performs well, but the result is slightly worse because it is forced to give more weight to the effect of propagating differences. The comparatively bad performance of the third algorithm seems to confirm this: it is designed to find propagating differences that affect groups of elements which do not seem to have a strong effect on the overall number of differences.
Also, since there seems to be no or at best little correlation between differences and element density, one can assume that the number of true errors made by the engravers is low and not affected by the density of elements.

Especially the example of the 5\textsuperscript{th} Bagatelle illustrates these observations. For all the comparison pairs and all Bagatelles, the results look like a norm distribution skewed to the right, but the fifth Bagatelle shows that this skewness is produced by the mentioned superimposition. Most of the differences are randomly distributed, which is the main part of the distributions. Added to this are few differences induced by propagating, deliberate choices. These are indeed correlated with the element density and thus have only an effect if bars containing the affected elements are selected for the sample. Only in the fifth Bagatelle this effect is strong enough to be visually represented in the results because the two bars heavily increase the number of differences (see section \ref{sec:results} and Figure \ref{fig:ex_5}). On the other hand, this example shows, how small the effect of propagating differences is. In bars 63 and 64 there are many propagating differences. Hence, in this case it would be likely that the \textit{barElCount} algorithm performs better than the \textit{randSel} algorithm. As can be seen in Table \ref{tab:results} this is not the case, the deviation of mean is larger for the \textit{barElCount} algorithm.

In other words: most differences between two editions of the same work affect only single elements, rarely a whole group of elements. This is not to say that the impact of these differences on performances is small, too. But in a graphical sense they are.

However, there are some limitations to our results. First, some of the differences occur because of the encoding. There are cases in which the position of an element is arbitrary. For example, the exact starting and end point of hairpins can be unclear because the two voices are not perfectly aligned. Also, for example the difference between staccato dots and wedges is not obvious: how ``long" may a dot be before it becomes a wedge? And, of course, there can be errors made in the encoding.

Second, we only used a very limited set of musical works for our research. A next step therefore has to be to extend the corpus of works under investigation. 
Third, we only considered six different editions of the works. This number should also be extended in future work. 
Another crucial limitation is that we used all pre-settings of the \textsf{musicdiff}-Tool. Especially, we did not change the costs for edit operations. The costs of each edit operation heavily influence the results of alignment algorithms. Changing costs based on musicological considerations can change the outcome of the investigation (see \cite{Merkl2022}). 
We also did not give more or less weight to special differences, but counted all differences as one difference. In the tradition of stemmatology, one could consider defining some kind of difference as ``more important'' than others. To do so, it would be necessary to analyze the results of the comparisons not only quantitatively but also qualitatively. Still, since we are first and foremost interested in a quantitative analysis, our findings allow us to conduct the desired research. Also, the distribution of differences is not influenced by musicological evaluation of their meaning. This is to say, that the position of differences does not change when they are qualitatively analyzed.

Despite these limitations, we consider our research to be a first step toward understanding editorial practice in 19\textsuperscript{th}-century German-speaking countries. The results lead us to discard our initial hypothesis, which assumed a correlation between element and edit operation distribution, stemming from the notion of deliberate editorial decisions. This notion does not seem to hold up anymore. As discussed, further investigation on a larger corpus is needed.

In addition to our own question regarding editorial practice, we believe that this approach is promising for other corpus studies as well, even though the development of sampling methods for symbolic music corpora is just at its beginning.
We showed that the analysis of a randomized subset of data can be used to conduct a meaningful analysis of musicological data. This finding shows that the use of sampling methods may provide in the future a possibility to reduce the workload/effort of preparing music encodings significantly. This will enable research not dependent on already existing corpora, so musicological questions do not have to match these corpora but the corpora can match the questions.

\bibliographystyle{apalike}
\bibliography{bibliography}

\begin{thebibliography}{}

\bibitem[Beer, 2000]{Beer2000}
Beer, A. (2000).
\newblock {\em Musik zwischen Komponist, Verlag und Publikum. Die
  Rahmenbedingungen des Musikschaffens in Deutschland im ersten Drittel des 19.
  Jahrhunderts}.
\newblock Verlag Hans Schneider, Tutzing.

\bibitem[Beer, 2020]{Beer2020}
Beer, A. (2020).
\newblock {\em Das Leipziger Bureau de Musique (Hoffmeister \& Kühnel, A.
  Kühnel). Geschichte und Verlagsproduktion (1800—1814)}.
\newblock Musikwissenschaftliche Schriften, Bd. 55. Musikverlag Bernd
  Katzbichler, München, Salzburg.

\bibitem[Burgoyne et~al., 2011]{burgoyne2011}
Burgoyne, J.~A., Wild, J., and Fujinaga, I. (2011).
\newblock An expert ground truth set for audio chord recognition and music
  analysis.
\newblock In {\em Proceedings of the 12th International Society for Music
  Information Retrieval Conference}, page 633—638.

\bibitem[Calvo-Zaragoza et~al., 2023]{calvo-zaragoza-2023}
Calvo-Zaragoza, J., Martinez-Sevilla, J.~C., Penarrubia, C., and Rios-Vila, A.
  (2023).
\newblock Optical music recognition: Recent advances, current challenges,
  and future directions.
\newblock In Coustaty, M. and Forn{\'e}s, A., editors, {\em Document Analysis
  and Recognition -- ICDAR 2023 Workshops}, pages 94--104, Cham. Springer
  Nature Switzerland.

\bibitem[Devaney et~al., 2015]{devaney2015}
Devaney, J., Arthur, C., Condit-Schultz, N., and Nisula, K. (2015).
\newblock Theme and variation encodings with roman numerals (tavern): A new
  data set for symbolic music analysis.
\newblock In {\em Proceedings of the 16th International Society for Music
  Information Retrieval Conference}, pages 728--734. ISMIR.

\bibitem[Foscarin et~al., 2019]{Foscarin2019}
Foscarin, F., Fournier-S'Niehotta, R., and Jacquemard, F. (November 2019).
\newblock A diff procedure for music score files.
\newblock In {\em Computation and visualization of the differences between two
  music score files. 6th International Conference on Digital Libraries for
  Musicology (DLfM)}, pages 7--13, Den Haag.

\bibitem[Goebl, 2021]{Goebl2021_MEIEncodingSieben}
Goebl, W. (2021).
\newblock {{MEI}} encoding: {{Sieben Bagatellen}}, {{Opus}} 33 by {{Ludwig}}
  van {{Beethoven}}.

\bibitem[Goehr, 1994]{Goehr1994}
Goehr, L. (1994).
\newblock {\em The imaginary museum of Musical Works}.
\newblock Oxford.

\bibitem[Hentschel et~al., 2025]{hentschel2025}
Hentschel, J., Rammos, Y., Neuwirth, M., and Rohrmeier, M. (2025).
\newblock A corpus and a modular infrastructure for the empirical study of
  (an)notated music.
\newblock {\em Scientific Data}, 12.

\bibitem[Lewis and Page, 2024]{Lewis2024}
Lewis, D. and Page, K. (2024).
\newblock Popular musical arrangements in the nineteenth-century home: A study
  of the harmonicon supported by digital tools.
\newblock In {\em Proceedings of the 11th International Conference on Digital
  Libraries for Musicology}, DLfM '24, page 32–39, New York. Association for
  Computing Machinery.

\bibitem[Lewis et~al., 2023]{Lewis2023}
Lewis, D., Shibata, E., Hankinson, A., Kepper, J., Page, K., Rosendahl, L.,
  Saccomano, M., and Siegert, C. (2023).
\newblock Supporting musicological investigations with information retrieval
  tools: an iterative approach to data collection.
\newblock In {\em Proceedings of the 24th International Society for Music
  Information Retrieval Conference}, pages 795--801, Milan, Italy.

\bibitem[London, 2013]{London2013}
London, J. (2013).
\newblock Building a representative corpus of classical music.
\newblock {\em Music Perception}, 31.

\bibitem[Mayer et~al., 2024]{Mayer2024}
Mayer, J., Straka, M., Haji{\v{c}}, J., and Pecina, P. (2024).
\newblock Practical end-to-end optical music recognition for pianoform music.
\newblock In Barney~Smith, E.~H., Liwicki, M., and Peng, L., editors, {\em
  Document Analysis and Recognition - ICDAR 2024}, pages 55--73, Cham. Springer
  Nature Switzerland.

\bibitem[Merkl, 2022]{Merkl2022}
Merkl, R. (2022).
\newblock {\em Bioinformatik. Grundlagen, Algorithmen, Anwendungen}.
\newblock Wiley-VCH, Weinheim, 4 edition.

\bibitem[Neuwirth et~al., 2018]{neuwirth2018}
Neuwirth, M., Harasim, D., Moss, F.~C., and Rohrmeier, M. (2018).
\newblock The annotated beethoven corpus (abc): A dataset of harmonic analyses
  of all beethoven string quartets.
\newblock {\em Frontiers in Digital Humanities}, 5.

\bibitem[R{\'i}os-Vila et~al., 2024]{Rios-Vila2024}
R{\'i}os-Vila, A., Calvo-Zaragoza, J., and Paquet, T. (2024).
\newblock Sheet music transformer: End-to-end optical music recognition beyond
  monophonic transcription.
\newblock In Barney~Smith, E.~H., Liwicki, M., and Peng, L., editors, {\em
  Document Analysis and Recognition - ICDAR 2024}, pages 20--37, Cham. Springer
  Nature Switzerland.

\bibitem[R{\'i}os-Vila et~al., 2023]{Rios-Vila2023}
R{\'i}os-Vila, A., Rizo, D., I{\~{n}}esta, J.~M., and Calvo-Zaragoza, J.
  (2023).
\newblock End-to-end optical music recognition for pianoform sheet music.
\newblock {\em International Journal on Document Analysis and Recognition
  (IJDAR)}.

\bibitem[Saccomano et~al., 2024]{Saccomano2024}
Saccomano, M., Rosendahl, L., Lewis, D., Hankinson, A., Kepper, J., Page, K.,
  and Shibata, E. (2024).
\newblock Selective encoding: Reducing the burden of transcription for digital
  musicologists.
\newblock In {\em Journal of the Text Encoding Initiative [online]}.

\bibitem[Shea et~al., 2024]{Shea2024}
Shea, N., Reymore, L., White, C.~W., Duinker, B., VanHandel, L., Zeller, M.,
  and Biamonte, N. (2024).
\newblock Diversity in music corpus studies.
\newblock {\em Music Theory Online}, 30(1 (February)).

\end{thebibliography}
\clearpage

\section{Tables}

\begin{table}[h]
\caption{Overview of the results for Bagatelle No. 1}
\label{tab:full_results-1}
\centering
\begin{tabular}{rrrrr}
\toprule
\#Comp & $\mu_{e,e'}$ & $\bar{\delta}_{randSel}$ & $\bar{\delta}_{barElCount}$ & $\bar{\delta}_{onlyEl}$\\ 
\midrule
\begin{filecontents*}{results_corr_rounded.csv}
Bagatelle, #Vgl., diff_full, µ_theo, µ_elem, std_elem,Abw theo elem, µ_mue, std_mue,Abw theo mue, µ_rm, std_rm,Abw theo rm
1.0,1.0,210.0,21.0,18.0,7.09,3.0,18.93,7.43,2.07,18.83,7.7,2.17
,2.0,277.0,27.7,24.81,8.02,2.89,25.94,8.05,1.76,25.95,8.1,1.75
,3.0,259.0,25.9,22.94,8.16,2.96,23.98,8.51,1.92,23.87,8.6,2.03
,4.0,384.0,38.4,36.61,9.63,1.79,38.17,9.63,0.23,38.17,9.99,0.23
,5.0,371.0,37.1,34.07,9.43,3.03,35.69,9.61,1.41,35.59,10.41,1.51
,6.0,267.0,26.7,24.15,7.62,2.55,25.12,7.75,1.58,25.14,7.88,1.56
,7.0,182.0,18.2,17.2,7.06,1.0,18.11,7.38,0.09,17.99,7.29,0.21
,8.0,398.0,39.8,35.82,10.26,3.98,37.48,10.58,2.32,37.52,10.98,2.28
,9.0,416.0,41.6,36.44,10.73,5.16,38.43,11.2,3.17,38.19,11.98,3.41
,10.0,316.0,31.6,29.02,8.85,2.58,30.34,9.0,1.26,30.29,9.1,1.31
,11.0,420.0,42.0,37.74,10.71,4.26,39.64,10.91,2.36,39.52,11.4,2.48
,12.0,465.0,46.5,42.13,11.63,4.37,44.14,11.82,2.36,44.04,12.12,2.46
,13.0,383.0,38.3,34.75,10.45,3.55,36.35,10.61,1.95,36.39,10.97,1.91
,14.0,430.0,43.0,38.29,11.26,4.71,40.25,11.63,2.75,40.06,12.21,2.94
,15.0,299.0,29.9,26.0,7.72,3.9,27.02,7.81,2.88,26.94,7.86,2.96
,,,0.0,31.43,9.24,3.32,31.97,9.46,1.87,31.9,9.62,1.95
2.0,1.0,227.0,22.7,18.44,6.43,4.26,19.57,6.37,3.13,19.83,6.32,2.87
,2.0,427.0,42.7,39.36,8.72,3.34,41.8,8.16,0.9,41.82,8.12,0.88
,3.0,445.0,44.5,39.19,9.26,5.31,41.65,8.36,2.85,41.84,8.41,2.66
,4.0,636.0,63.6,53.79,12.21,9.81,57.21,10.6,6.39,57.17,10.61,6.43
,5.0,538.0,53.8,48.52,11.3,5.28,51.37,10.11,2.43,51.58,10.23,2.22
,6.0,382.0,38.2,34.29,7.82,3.91,36.5,7.54,1.7,36.53,7.5,1.67
,7.0,364.0,36.4,32.97,7.86,3.43,35.11,6.96,1.29,35.18,6.98,1.22
,8.0,588.0,58.8,49.56,11.38,9.24,52.66,10.17,6.14,52.63,10.22,6.17
,9.0,510.0,51.0,46.83,10.96,4.17,49.53,9.95,1.47,49.67,10.07,1.33
,10.0,388.0,38.8,36.34,8.72,2.46,38.61,8.34,0.19,38.77,8.39,0.03
,11.0,503.0,50.3,42.47,11.27,7.83,45.18,10.42,5.12,45.09,10.62,5.21
,12.0,418.0,41.8,39.05,9.49,2.75,41.3,8.84,0.5,41.32,8.96,0.48
,13.0,431.0,43.1,36.54,9.6,6.56,38.79,9.05,4.31,38.87,9.1,4.23
,14.0,485.0,48.5,44.67,10.19,3.83,47.31,9.29,1.19,47.48,9.42,1.02
,15.0,424.0,42.4,36.67,9.99,5.73,38.78,9.48,3.62,38.8,9.58,3.6
,,,0.0,39.91,9.68,5.19,42.36,8.91,2.75,42.44,8.97,2.67
3.0,1.0,153.0,15.3,13.15,3.83,2.15,13.62,3.79,1.68,13.77,3.91,1.53
,2.0,453.0,45.3,42.35,7.91,2.95,44.58,8.07,0.72,44.53,8.3,0.77
,3.0,448.0,44.8,41.44,7.17,3.36,43.52,7.3,1.28,43.58,7.85,1.22
,4.0,604.0,60.4,57.99,12.89,2.41,61.22,13.27,0.82,61.01,13.73,0.61
,5.0,618.0,61.8,58.81,9.9,2.99,61.83,10.05,0.03,61.72,10.84,0.08
,6.0,433.0,43.3,40.98,7.54,2.32,43.1,7.69,0.2,43.01,8.0,0.29
,7.0,404.0,40.4,37.66,6.72,2.74,39.56,6.89,0.84,39.59,7.41,0.81
,8.0,388.0,38.8,53.61,12.52,14.81,56.68,12.92,17.88,56.38,13.29,17.58
,9.0,390.0,39.0,54.95,9.82,15.95,57.78,9.94,18.78,57.71,10.6,18.71
,10.0,186.0,18.6,14.92,4.87,3.68,15.62,4.87,2.98,15.6,4.87,3.0
,11.0,334.0,33.4,30.74,10.23,2.66,32.59,10.47,0.81,32.28,10.41,1.12
,12.0,351.0,35.1,31.98,8.81,3.12,33.53,8.82,1.57,33.48,8.95,1.62
,13.0,255.0,25.5,24.14,9.82,1.36,25.6,10.08,0.1,25.46,10.09,0.04
,14.0,292.0,29.2,27.49,8.07,1.71,28.92,8.15,0.28,28.84,8.42,0.36
,15.0,392.0,39.2,37.83,14.96,1.37,40.0,15.25,0.8,39.8,15.32,0.6
,,,0.0,37.87,9.0,4.24,39.88,9.17,3.25,39.78,9.47,3.22
4.0,1.0,269.0,26.9,25.1,6.8,1.8,25.4,6.96,1.5,25.26,7.39,1.64
,2.0,309.0,30.9,26.06,6.78,4.84,26.39,6.9,4.51,26.26,6.98,4.64
,3.0,326.0,32.6,28.67,8.1,3.93,29.09,8.18,3.51,28.93,8.3,3.67
,4.0,408.0,40.8,36.54,9.5,4.26,37.0,9.59,3.8,36.83,9.63,3.97
,5.0,412.0,41.2,37.24,8.89,3.96,37.96,8.59,3.24,37.65,8.53,3.55
,6.0,231.0,23.1,20.26,4.84,2.84,20.54,4.79,2.56,20.39,4.95,2.71
,7.0,213.0,21.3,18.55,6.49,2.75,18.86,6.53,2.44,18.74,6.64,2.56
,8.0,380.0,38.0,29.07,7.34,8.93,29.49,7.29,8.51,29.27,7.34,8.73
,9.0,459.0,45.9,41.76,8.77,4.14,42.44,8.68,3.46,42.13,8.72,3.77
,10.0,258.0,25.8,20.98,6.45,4.82,21.29,6.17,4.51,21.06,6.14,4.74
,11.0,320.0,32.0,27.22,7.56,4.78,27.62,7.22,4.38,27.37,7.17,4.63
,12.0,448.0,44.8,42.77,8.6,2.03,43.45,8.23,1.35,43.17,8.21,1.63
,13.0,238.0,23.8,19.83,7.07,3.97,20.09,6.72,3.71,19.86,6.73,3.94
,14.0,427.0,42.7,36.96,9.35,5.74,37.54,8.9,5.16,37.27,8.9,5.43
,15.0,375.0,37.5,34.04,8.42,3.46,34.43,8.03,3.07,34.27,7.94,3.23
,,,0.0,29.67,7.67,4.15,30.11,7.52,3.71,29.9,7.57,3.92
5.0,1.0,178.0,17.8,18.01,8.07,0.21,19.69,8.45,1.89,19.98,8.92,2.18
,2.0,364.0,36.4,31.87,10.69,4.53,35.49,10.0,0.91,35.49,10.07,0.91
,3.0,365.0,36.5,32.42,8.36,4.08,35.83,8.29,0.67,36.01,8.47,0.49
,4.0,424.0,42.4,42.43,12.08,0.03,47.11,11.85,4.71,46.77,12.08,4.37
,5.0,493.0,49.3,46.25,24.61,3.05,51.62,25.32,2.32,51.19,25.44,1.89
,6.0,382.0,38.2,27.31,9.51,10.89,30.44,8.32,7.76,30.27,8.4,7.93
,7.0,292.0,29.2,27.37,7.88,1.83,30.23,7.99,1.03,30.44,8.06,1.24
,8.0,400.0,40.0,39.37,11.73,0.63,43.76,11.63,3.76,43.39,11.95,3.39
,9.0,475.0,47.5,44.7,25.69,2.8,50.44,25.54,2.94,49.65,25.6,2.15
,10.0,341.0,34.1,28.9,8.88,5.2,32.1,8.47,2.0,32.11,8.61,1.99
,11.0,396.0,39.6,36.74,13.19,2.86,40.92,12.64,1.32,40.51,12.53,0.91
,12.0,496.0,49.6,45.23,22.3,4.37,50.72,22.53,1.12,50.05,22.62,0.45
,13.0,295.0,29.5,29.83,12.98,0.33,33.39,13.25,3.89,33.1,13.22,3.6
,14.0,522.0,52.2,49.08,24.2,3.12,54.77,24.9,2.57,54.37,25.11,2.17
,15.0,478.0,47.8,47.4,24.46,0.4,52.88,25.26,5.08,52.24,25.36,4.44
,,,39.34,36.46,14.97,2.96,40.63,14.96,2.8,40.37,14.36,2.54
\end{filecontents*}
\csvreader[
head to column names,
table head={\hline \textbf{Spalte 1} & \textbf{Spalte 2} & \textbf{Spalte 3} & \textbf{Spalte 4} & \textbf{Spalte 5} & \textbf{Spalte 6} \\ \hline},
respect all,
late after line=\\,
range = {1-15}
]{results_corr_rounded.csv}{1=\csvcoli,2=\csvcolii}{
\csvcolii & \csvcoliv & \csvcolxi & \csvcolviii & \csvcolv
}
\bottomrule
\end{tabular}
\end{table}

\begin{table}[h]
\caption{Overview of the results for Bagatelle No. 2}
\label{tab:full_results-2}
\centering
\begin{tabular}{rrrrr}
\toprule
\#Comp& $\mu_{e,e'}$ & $\bar{\delta}_{randSel}$ & $\bar{\delta}_{barElCount}$ & $\bar{\delta}_{onlyEl}$\\ 
\midrule
\csvreader[
head to column names,
table head=\hline \textbf{Spalte 1} & \textbf{Spalte 2} & \textbf{Spalte 3} & \textbf{Spalte 4} & \textbf{Spalte 5} & \textbf{Spalte 6} \\ \hline,
respect all,
late after line=\\,
range = {17-31}
]{results_corr_rounded.csv}{1=\csvcoli,2=\csvcolii}{
\csvcolii & \csvcoliv & \csvcolxi & \csvcolviii & \csvcolv
}
\bottomrule
\end{tabular}
\end{table}

\begin{table}[h]
\caption{Overview of the results for Bagatelle No. 3}
\label{tab:full_results-3}
\centering
\begin{tabular}{rrrrr}
\toprule
\#Comp & $\mu_{e,e'}$ & $\bar{\delta}_{randSel}$ & $\bar{\delta}_{barElCount}$ & $\bar{\delta}_{onlyEl}$\\
\midrule
\csvreader[
head to column names,
table head=\hline \textbf{Spalte 1} & \textbf{Spalte 2} & \textbf{Spalte 3} & \textbf{Spalte 4} & \textbf{Spalte 5} & \textbf{Spalte 6} \\ \hline,
respect all,
late after line=\\,
range = {33-47}
]{results_corr_rounded.csv}{1=\csvcoli,2=\csvcolii}{
\csvcolii  & \csvcoliv & \csvcolxi & \csvcolviii & \csvcolv
}
\bottomrule
\end{tabular}
\end{table}

\begin{table}[h]
\caption{Overview of the results for Bagatelle No. 4}
\label{tab:full_results-4}
\centering
\begin{tabular}{rrrrr}
\toprule
\#Comp& $\mu_{e,e'}$ & $\bar{\delta}_{randSel}$ & $\bar{\delta}_{barElCount}$ & $\bar{\delta}_{onlyEl}$\\ 
\midrule
\csvreader[
head to column names,
table head=\hline \textbf{Spalte 1} & \textbf{Spalte 2} & \textbf{Spalte 3} & \textbf{Spalte 4} & \textbf{Spalte 5} & \textbf{Spalte 6} \\ \hline,
respect all,
late after line=\\,
range = {49-63}
]{results_corr_rounded.csv}{1=\csvcoli,2=\csvcolii}{
\csvcolii & \csvcoliv & \csvcolxi & \csvcolviii & \csvcolv
}
\bottomrule
\end{tabular}
\end{table}

\begin{table}[h]
\caption{Overview of the results for Bagatelle No. 5}
\label{tab:full_results-5}
\centering
\begin{tabular}{rrrrr}
\toprule
\#Comp&$\mu_{e,e'}$&$\bar{\delta}_{randSel}$&$\bar{\delta}_{barElCount}$&$\bar{\delta}_{onlyEl}$\\ 
\midrule
\csvreader[
head to column names,
table head=\hline \textbf{Spalte 1} & \textbf{Spalte 2} & \textbf{Spalte 3} & \textbf{Spalte 4} & \textbf{Spalte 5} & \textbf{Spalte 6} \\ \hline,
respect all,
late after line=\\,
range = {65-79}
]{results_corr_rounded.csv}{1=\csvcoli,2=\csvcolii}{
\csvcolii & \csvcoliv & \csvcolxi & \csvcolviii & \csvcolv
}
\bottomrule
\end{tabular}
\end{table}
\cleardoublepage

\section{Algorithms}

\begin{algorithm}[h]
\caption{Random selection}
\label{algo:random}
\begin{algorithmic}[1]
\Procedure{Random selection}{}

random\_selection(measures, sample\_size = None)

\textit{Check, if variable sample\_size was given by user:}

\If{sample\_size is not None}
	\If {sample\_size $>$ number of measures}
		\State Error
	\Else
		\State required\_measures = sample\_size
	\EndIf

\Else

	\State required\_measures = ROUND UP(10\% of the number of measures)
\EndIf

\textit{Declare variable valid\_measures:}

\State valid\_measures = List of keys in the ``measures''-Dictionary

\textit{Select measures:}

\State selected\_measures = random selection of \#required\_measures from List valid\_measures without replacement

\Return selected\_measures 

END
\EndProcedure
\end{algorithmic}
\end{algorithm}

\begin{algorithm}[h]
\caption{Selection based on number of measures and elements.}
\label{algo:mae}
\begin{algorithmic}[1]
\Procedure{Selection of bars and elements}{}

\textit{Define function measures\_elements taking measures and sample\_size as arguments. Default value of sample\_size set to None}

\State measures\_elements(measures, sample\_size = None)

\textit{Declare lists measure\_selection, selected\_elements and available\_measures}

\State measure\_selection = empty\_list
\State selected\_elements = empty\_list
\State available\_measures = List of keys in the ``measures''-Dictionary

\textit{Check, if variable sample\_size was given by user:}

\If{sample\_size is not None:}

	\If {sample\_size $>$ number of measures}
		\State Error
	\Else
		\State required\_measures = sample\_size
	\EndIf

\Else

	\State required\_measures = ROUND\_UP(10\% of the number of measures)
\EndIf

\textit{Define percentage share of required\_measures of all measures}

\State measure\_ratio = DIVIDE required\_measures by number\_of\_all\_measures

\textit{Define number of required elements based on measure\_ratio}

\State required\_elements = Number of all elements * measure\_ratio

\textit{Declare variable max\_attempts, so that the algorithm stops at some point if no solution can be found}

\State max\_attempts = Number of measures
\State attempts = 0

\textit{Set boundaries for accepted results:}

\State lower\_bound = 95\% of required\_elements

\State upper\_bound = 105\% of required\_elements

\textit{Declare variable current\_sum:}

\State current\_sum = sum of elements in the selected measures

\textit{Select measures as long as number of required\_measures isn't reached or number of elements is smaller than lower\_bound:}

\While {length of measure\_selection $\neq$ required\_measures OR current\_sum $<$ lower\_bound} 
	\State selected\_measures = add random selection out of List available\_measures
	\State add number of elements in selected measure to selected\_elements
	\State update current\_sum
	
	\textit{Remove measures from measure\_selection if number of required\_measures is exceeded}
	\If {length of measure\_selection $>$ required\_measures}
		\State randomly remove one measure from measure\_selection
		\State remove the number of elements in this bar from selected\_elements
		\State update current\_sum 
	\EndIf
		\algstore{myalg1}
\end{algorithmic}
\end{algorithm}
\begin{algorithm}[h]
\ContinuedFloat
\caption{Selection based on number of measures and elements (continued)}
\begin{algorithmic}
\algrestore{myalg1}
	\State raise attempts by 1
	
	\textit{If upper\_bound is exceeded, remove measures until current\_sum $<$ upper\_bound}
	\While {current\_sum $>$ upper\_bound}
		\State randomly remove one measure from measure\_selection
		\State remove the number of elements in this bar from selected\_elements
		\State update current\_sum 
	\EndWhile

	\If{attempts == max\_attempts}
		\State break
	\EndIf
\EndWhile
\If {current\_sum $<$ lower\_bound}
	\State print warning
\ElsIf {current\_sum $>$ upper\_bound}
	\State print warning
\Else
	\State print celebration
\EndIf

\Return selected\_measures 

\EndProcedure
\end{algorithmic}
\end{algorithm}

\begin{algorithm}[h]
\caption{Selection based only on element distribution.}
\label{algo:elem}
\begin{algorithmic}[1]
\Procedure{Selection of bars only based on number of elements}{}

only\_elements(measures)

\textit{Declare lists measure\_selection, selected\_elements and available\_measures}

\State measure\_selection = empty\_list
\State selected\_elements = empty\_list
\State available\_measures = List of keys in the ``measures''-Dictionary

\textit{Define number of required elements}

\State required\_elements = ROUND\_UP(10\% of all elements in the piece)

\textit{Declare variable max\_attempts, so that the algorithm stops at some point if no solution can be found:}

\State max\_attempts = Number of measures
\State attempts = 0

\textit{Set boundaries for accepted results:}

\State lower\_bound = 95\% of required\_elements

\State upper\_bound = 105\% of required\_elements

\textit{Declare variable current\_sum:}

\State current\_sum = sum of elements in the selected measures

        \textit{Choose measures until current\_sum reaches lower\_bound or attempts $>$ max\_attempts}
        \While{current\_sum $<$ lower\_bound AND attempts $<$ max\_attempts} 
            	\State x = randomly chosen measure from available\_measures
            	\State ADD x to measure\_selection
            	\State ADD number of elements in x to selected\_elements
		\State REMOVE x from available\_measures
            	\State UPDATE current\_sum
            	\State attempts += 1
        \EndWhile
        
        \textit{Randomly remove measures from the selection if current\_sum is larger than upper\_bound}
                \While {current\_sum $>$ upper\_bound}
                		\State r =  randomly chosen measure from measure\_selection
			\State REMOVE r from selected\_measures
        	        		\State REMOVE number of elements in r from selected\_elements
                		\State UPDATE current\_sum
                \EndWhile

            \If{attempts == max\_attempts}
            	\State RETURN SORTED(measure\_selection)
            \EndIf
        \If{current\_sum $<$ lower\_bound OR current\_sum $>$ upper\_bound}
        	\State PRINT warning
        
        \Else
        
        	\Return sorted measure\_selection
        \EndIf

\EndProcedure
\end{algorithmic}
\end{algorithm}

\end{document}